\documentclass[aip,jcp,amsmath,amssymb,reprint]{revtex4-2}

\usepackage{braket}
\usepackage{hyperref}
\usepackage{xcolor}
\usepackage{graphicx}

\definecolor{BrickRed}{cmyk}{0, .89, .94, .28}

\newcommand{\be}{\begin{equation}}
\newcommand{\ee}{\end{equation}}
\newcommand{\vs}{v_{\rm S}}
\newcommand{\dvs}{\dot{v}_{\rm S}}

\newcommand{\vext}{v_{\rm ext}}
\newcommand{\dvext}{\dot{v}_{\rm ext}}
\newcommand{\vhxc}{v_{\rm Hxc}}

\newcommand{\inte}{\int \!}
\newcommand{\intdr}{\inte d{\bf r}\;}
\newcommand{\intdrr}{\inte d{\bf r}'\;}
\newcommand{\rt}{({\bf r},t)}
\newcommand{\rrt}{({\bf r}',t)}
\newcommand{\ddn}{\ddot{n}}
\newcommand{\Vr}{\mathcal{V}_{({\bf r})}}

\newcommand{\bfr}{{\bf r}}

\usepackage{stackengine}
\newcommand\fit[3][.3ex]{\stackengine{#1}{#3}{#2}{O}{c}{F}{T}{S}}
\newcommand\umlaut[1]{\fit{\kern-.05ex.\kern-.15ex.}{#1}}
\begin{document}

\title{A reformulation of time-dependent Kohn-Sham theory in terms of the second time derivative of the density}

\author{Walter Tarantino}
\email{walter.tarantino@dsf.unica.it}
\affiliation{Dipartimento di Fisica, Università degli Studi di Cagliari, Cittadella Universitaria, I-09042 Monserrato, Cagliari, Italy}
\author{Carsten A. Ullrich}
\affiliation{Department of Physics and Astronomy, University of Missouri, Columbia, Missouri 65211, USA}
\date{\today}

\begin{abstract}
The Kohn-Sham approach to time-dependent density-functional theory (TDDFT) can be formulated, in principle exactly, by invoking the force-balance equation for the density,
which leads to an explicit expression for the exchange-correlation potential as an implicit density functional. It is shown that this suggests a
reformulation of TDDFT in terms of the second time derivative of the density, rather than the density itself. The result is a time-local Kohn-Sham scheme
of second order in time whose causal structure is more transparent than that of the usual Kohn-Sham formalism. The scheme can be used to construct
new approximations at the exchange-only level and beyond, and it offers a straightforward definition of the exact adiabatic approximation.
\end{abstract}

\maketitle

\section{Introduction}

Over the last decades, time-dependent density functional theory
(TDDFT) has become the method of choice for calculating
dynamical properties of electronic many-body systems. \cite{Ullrich2012,Maitra2016}
The basic theorem of TDDFT due to Runge and Gross (RG) \cite{Runge1984} establishes the one-to-one
correspondence between time-dependent densities and one-body potentials, for given initial states.
Thus, in spite of some subtle open questions regarding the existence and uniqueness of the density-potential mapping,
\cite{Yang2012,Yang2013,Ruggenthaler2011,Ruggenthaler2012,Ruggenthaler2015}
TDDFT represents a complete and formally exact reformulation of the quantum mechanical many-body problem, whose
central quantity is the electronic density $n(\bfr,t)$.

The many-body Schr\"odinger equation is a conceptually straightforward initial value problem:
start with an initial many-body state and propagate it forward under the influence of a given time-dependent
external potential. However, the complexity of the wave function increases exponentially with the number of
particles. \cite{Kohn1999} Using the density as basic variable is therefore a very attractive
idea, recognizing that the many-body wave function is not necessary to describe the properties of interacting electrons.
In this paper, we closely examine this idea and its implications, and then offer a new perspective.

We first ask the following question: TDDFT tells us that all observables can be expressed as functionals
of the density, but what do these functionals actually look like? At the formally exact level, a density
functional can be represented as a procedure, where the many-body wave function is ``hidden'' as an internal variable.
As a consequence, a density functional at a given time $t$ contains a memory which goes back all the way to the initial time.
This affects the nature of the initial value problem in TDDFT in a profound manner.

In the seminal work by van Leeuwen, \cite{vanLeeuwen1999} an explicit link between the Kohn-Sham (KS) and
the many-body system was established using an equation of motion for the density, often referred to as the force-balance equation.
This formalism has been used as the basis for fixed-point proofs,\cite{Ruggenthaler2011,Ruggenthaler2012,Ruggenthaler2015}
to demonstrate the causality of the time-dependent KS approach,\cite{Maitra2008}
and to study the density-potential mapping and for constructing approximations.\cite{Maitra2010,Nielsen2018,Tchenkoue2019,Lacombe2019,Brown2020}

We use the force-balance equation to formulate a closed prescription for the time propagation of
the exact KS-TDDFT scheme, which is written in the form of an initial value problem that is nonlocal in time, accounting for the inherent memory effects.
From this, we arrive at the central insight in this paper:
by using the second time derivative of the density, $\ddot n(\bfr,t)$ as basic variable instead of $n(\bfr,t)$, we
can formulate an alternative KS scheme that is effectively local in time, i.e., the propagation at time $t$ is entirely determined by (many-body)
objects evaluated at the same time $t$. We refer to this alternative scheme as KS-TD{\umlaut D}FT.

KS-TD{\umlaut D}FT still includes all the complexities of the many-body problem. However, it represents a
time-local initial value problem, which has conceptual and, possibly, practical advantages. In addition to discussing
the formally exact framework of KS-TD{\umlaut D}FT, we show that it is consistent with and can be reduced to standard KS theory.
We also show that it yields the known exchange-only approximation of TDDFT in the appropriate limit, and that it
provides fresh insight into the definition and meaning of the adiabatic approximation.

The paper is organized as follows. In Section \ref{secDF} we characterize density functionals as maps connecting the density to any given observable.
In Section \ref{secKSTDDFT}, standard KS-TDDFT is formulated using the force-balance equation.
We then make the transition to KS-TD{\umlaut D}FT
in Section \ref{secKSTD"DFT} and show that it can be realized in several variants.
In Section \ref{secAppr} we discuss two approximations: exact exchange and beyond, and the adiabatic approximation.
Conclusions are drawn in Section \ref{secConclusions},
while several appendices provide further technical details.

The following notation is used throughout the paper.
Italicized letters ($f$) are used for functions,
a hat ($\hat{f}$) denotes operators acting
on the Hilbert space of physical states,
and calligraphic letters ($\mathcal{F}$) are used for functionals
(i.e., maps between functions and/or quantum states).
This expresses the fact that a certain observable, say $O\rt$,
can be calculated either in the standard quantum mechanical way,
$O\rt=\braket{\hat{O}}$, or the density-functional way,
$O\rt=\mathcal{O}[n]$.
Furthermore, we distinguish between the value of a
function at a specific point in (space)time
[e.g., $n\rt$, $\vext\rt$, $\Psi(t)$]
and the set of \emph{all} the values over its domain
$\{ t_0\leq t<\infty,{\bf r}\in \mathbb{R}^3\}$,
denoted by the object without argument ($n$, $\vext$, $\Psi$).

In the following we will ignore the electronic spin.
When explicitly referring to initial times $t_0$, we will assume that $t_0=0$.
Atomic units ($\hbar=e=m=4\pi\epsilon_0 = 1$) will be used throughout this paper.

\section{Characterizing time-dependent density functionals}\label{secDF}

We begin by formulating the time-dependent density functional
for the generic observable $O(t)$, that is to say, a set of equations that completely characterizes the map
\be\label{Omap}
\mathcal{O}:(\Psi_0,n)\to O.
\ee
Let us start from the Schr\"odinger equation, i.e., the initial value problem
\be\label{inipsi}
\left\{\begin{array}{l}
i\partial_t\Psi(t)=\left(\hat{T}+\hat{W}+\intdr \vext \rt \hat{n}({\bf r}) \right)\Psi(t)\\
\Psi(0)=\Psi_0,
\end{array}\right.
\ee
where $\hat{T}$, $\hat{W}$, and  $\hat{n}$ are the operators of kinetic energy,
interaction and density, respectively.
This constitutes the definition of the functional $(\Psi_0,\vext)\to \Psi$:
given the initial state and the external potential
at all times, the set \eqref{inipsi} completely identifies the
wave function $\Psi(t)$ at all times.
One can then define the potential-functional of any observable via $O=\braket{\hat{O}}$ as follows:
\be\label{pfO}
\left\{\begin{array}{l}
O\rt=\bra{\psi(t)}\hat{O}({\bf r})\ket{\psi(t)}\\
\\
i\partial_t\psi(t)=\left(\hat{T}+\hat{W}+\intdr \vext \rt \hat{n}({\bf r}) \right)\psi(t)\\
\psi(0)=\Psi_0
\end{array}\right.
\ee
in which $\psi(t)$ denotes an internal variable,
as $O\rt$ is considered the sole output of the functional.
In the end, $\psi(t)$ is still equal to the many-body wave function $\Psi(t)$,
which is the output of the functional \eqref{inipsi},
but now it is hidden within the functional \eqref{pfO}.

The mapping $(\Psi_0,\vext)\to O$ is causal: the set (\ref{pfO}) shows that
$O\rt$ only depends on $\{v_{\rm ext}(0\leq t'< t)\}$ and, of course, the initial state $\Psi_0$.
In other words, to calculate the observable at time $t$ one only needs to know the
external potential in the past up until $t$, not in the present nor in the future.
It is important to note that the map itself only features
$\vext$ and $\psi$ at time $t$, and that this completely encodes the information
of $\vext$ at all previous times.
More precisely, to propagate $O\rt$ we do not need to always use all of  $\{v_{\rm ext}(0\leq t'\leq t)\}$,
but just $\psi(t)$ and $\vext(t)$ as we go along.
All the memory of the functional is carried within the wave function at time $t$;
the implementation is thus \emph{time-local}.

The RG theorem \cite{Runge1984} allows us
to turn the potential-functional into a density-functional:
\be\label{dfO}
\left\{\begin{array}{l}
O\rt=\bra{\psi(t)}\hat{O}({\bf r})\ket{\psi(t)}\\
\\
i\partial_t\psi(t)=\left(\hat{T}+\hat{W}+\intdr v \rt \hat{n}({\bf r}) \right)\psi(t)\\
\psi(0)=\Psi_0\\
\\
n\rt=\bra{\psi(t)} \hat{n}({\bf r}) \ket{\psi(t)}\\
\mbox{gauge condition on }v\rt.
\end{array}\right.
\ee
Here, the input is the density $n$ and the initial state $\Psi_0$,
the output is the observable $O$, and $\psi$ and $v$ are internal variables.
The RG theorem guarantees that, once a certain density is plugged in,
there is only one $v$ that satisfies the system of equations, namely the $\vext$
that gives rise the that very density.
In other words, the map  \eqref{Omap} can now be recast as
\be
(\Psi_0,n)\to(\Psi_0,v)\to \psi\to O.
\ee

While this provides a comprehensive formal definition
of the time-dependent density functional associated with a given observable,
it is not very useful for practical purposes:
to evaluate \eqref{dfO} for an actual situation one must not only
be able to solve the Schr\"odinger initial value problem, but also
to find the adequate $v$ that gives rise to the given $n$,
for instance, via a guess-and-check algorithm.

For an alternative, more convenient definition,
let us focus on the input of the functional.
For sufficiently smooth densities (as required by the RG theorem),
the information about $n$ is completely encoded by
\be
\left\{\begin{array}{l}
n({\bf r},0)\\
\partial_t n({\bf r},0)\\
\partial_t^2 n\rt.
\end{array}\right.
\ee
The second time-derivative of the density, and the associated initial conditions, can then be obtained as follows:
\be\label{7}
\left\{\begin{array}{l}
n({\bf r},0)=\bra{\Psi_0} \hat{n}({\bf r}) \ket{\Psi_0}\\
\partial_t n({\bf r},0)=-\bra{\Psi_0} \nabla\cdot\hat{{\bf j}}({\bf r}) \ket{\Psi_0}\\
\partial_t^2 n\rt=\nabla \cdot (
\bra{\psi(t)} \hat{n}({\bf r}) \ket{\psi(t)}
\nabla v \rt)\\
\hfill +\bra{\psi(t)} \hat{q}({\bf r}) \ket{\psi(t)},
\end{array}\right.
\ee
where $\hat{q} ({\bf r})=[\hat{T}+\hat{W},[\hat{T},\hat{n}({\bf r})]]
= \partial_\mu\partial_\nu\hat T_{\mu\nu}({\bf r}) + \partial_\mu \hat W_\mu({\bf r})$,
featuring the operators of the momentum-stress tensor $\hat T_{\mu\nu}({\bf r})$
and the divergence of the interaction-stress tensor $\hat W_\mu({\bf r})$.\cite{vanLeeuwen1999}

From the perspective of the functional \eqref{dfO},
the first two equations in (\ref{7}) are redundant,
for both $n$ and $\Psi_0$ are given and trivially satisfy those equations,
as long as a physical density is plugged in.
As a consequence of the RG theorem, the density-potential map
\be
(\Psi_0,n)\leftrightarrow (\Psi_0,\vext)
\ee
thus implies an alternative map of the form
\be
(\Psi_0,\ddot{n})\leftrightarrow (\Psi_0,\vext),
\ee
where $\ddot{n}$ denotes the set $\{\partial_t^2 n({\bf r},t\geq 0)\}$.
This has important implications for our definition of the density functional:
by rewriting the set \eqref{dfO} as
\be\label{ddfO}
\left\{\begin{array}{l}
O\rt=\bra{\psi(t)}\hat{O}({\bf r})\ket{\psi(t)}\\
\\
i\partial_t\psi(t)=\left(\hat{T}+\hat{W}+\intdr v \rt \hat{n}({\bf r}) \right)\psi(t)\\
\psi(0)=\Psi_0\\
\\
\partial_t^2 n\rt=\nabla \cdot (
\bra{\psi(t)} \hat{n}({\bf r}) \ket{\psi(t)}
\nabla v \rt)+\\
\hfill+\bra{\psi(t)} \hat{q}({\bf r}) \ket{\psi(t)}\\
\mbox{gauge condition on }v\rt,
\end{array}\right.
\ee
we can now use the last group of equations to write
$v$ at a given time $t$ in terms of $\ddot{n}$ and $\psi$
\emph{at the same time} $t$. We denote this by
\be\label{vV}
v\rt=\mathcal{V}_{(\bf r)}[\ddot{n}(t)-\braket{\hat{q}(t)},\braket{\hat{n}(t)}],
\ee
where the subscript ${\bf r}$ reminds us that the operator acts only on the coordinates
and not on the time variable,
and expectation values are taken with respect to $\psi(t)$.
Existence and computation of the operator $\mathcal{V}$ are
discussed in Appendix \ref{appV}.
Equation \eqref{vV} then allows us to rewrite the set \eqref{ddfO} in the following manner:
\be\label{ddVfO}
\left\{\begin{array}{l}
O\rt=\bra{\psi(t)}\hat{O}({\bf r})\ket{\psi(t)}\\
\\
i\partial_t\psi(t)=\Big(\hat{T}+\hat{W}+\\
\;\;\;\;\;+\intdr \mathcal{V}_{(\bf r)}[\ddot{n}(t)-\braket{\hat{q}(t)},\braket{\hat{n}(t)}]
    \hat{n}({\bf r}) \Big)\psi(t)\\
\psi(0)=\Psi_0.
\end{array}\right.
\ee
This system represents a complete characterization of the map \eqref{Omap}, alternative to \eqref{dfO}.
On one hand, the time-dependent many-body Schr\"odinger equation has turned into
a different problem, in which the external potential is replaced by a many-body object with a complex internal structure;
on the other hand, we no longer need an inversion algorithm to find the $v$ that determines $n$ (which, importantly, involves times prior to $t$).
As a consequence of this, we have arrived at a {\em time-local} initial value problem:
all quantities appearing on the right-hand side of the
Schr\"odinger equation in (\ref{ddVfO}) only depend on the time $t$
at which the derivative at the left-hand side is calculated---provided, however, that we look at $\ddn \rt$ as a given object (i.e., the input into
the functional), rather than the actual second time derivative of the density  (i.e., an object to be determined).

The above rewriting also clarifies the causal structure of the mapping,
for it shows that $O\rt$ depends only on $\{\ddot{n}(0\leq t'< t)\}$.
As before [see the remarks after Eq. (\ref{pfO})], to actually calculate $O\rt$ we do not need to keep track
of all past values of $\ddn$ as we go along, but $\psi(t)$
and $\ddn(t)$ at time $t$ are sufficient.
Recasting the problem in the form of Eq. (\ref{ddVfO}) therefore suggests that
adopting $\ddn\rt$ as fundamental quantity simplifies the structure of the theory.
In the following we shall take further advantage of this idea.

%%%%%%%%%%%%%%%%%%%%%%%%%%%%%%%%%%%%%%%%%%%%%%%%%%%%%%%%%%%%%%%%%%%%%%%%%%%%%%%%%%%%%%%%%%%%%%%%%
\begin{figure*}
\centering
\includegraphics[width=0.7\textwidth]{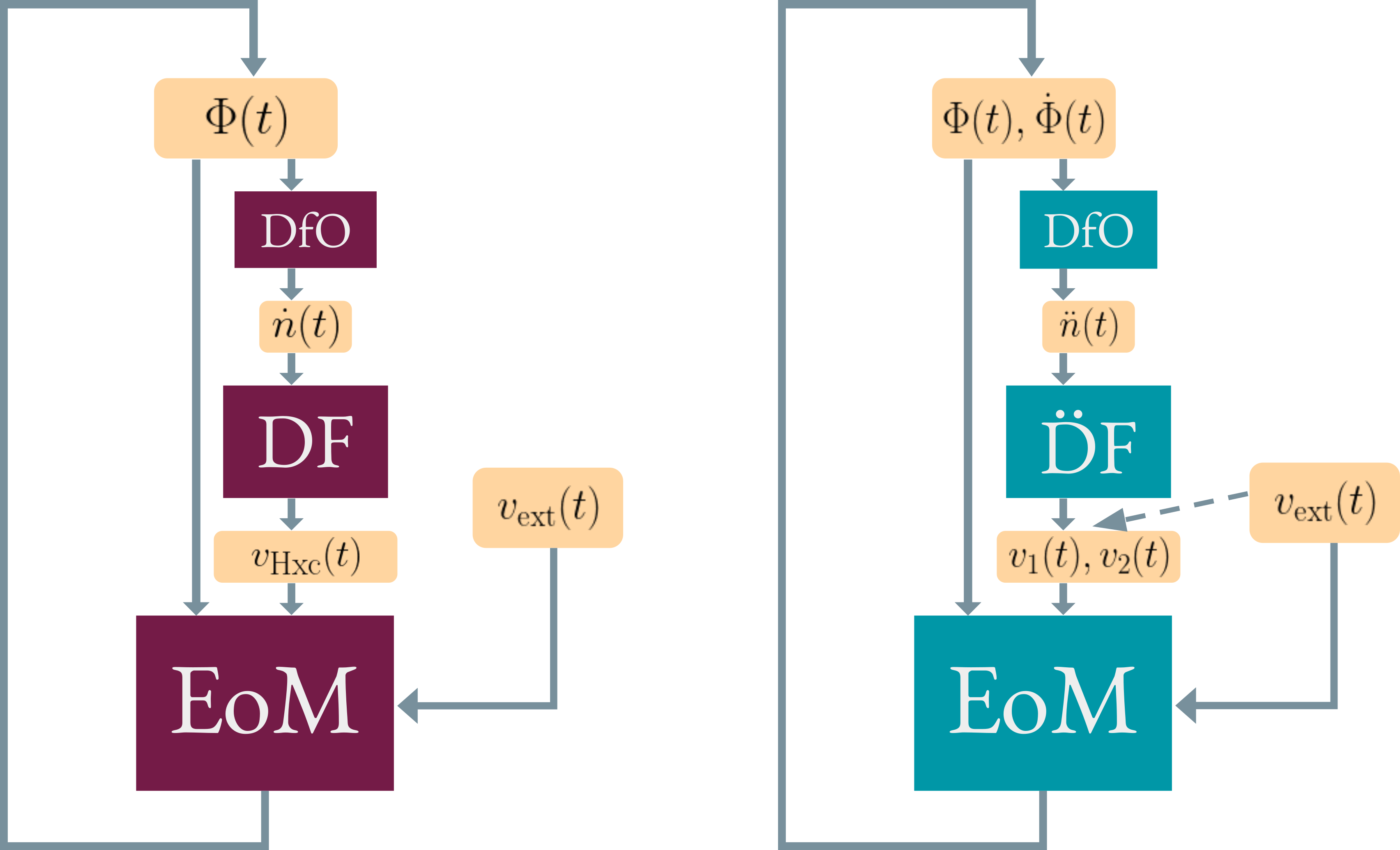}
\caption{Flowchart of the full set of equations for 
KS-TDDFT (left) and KS-TD{\umlaut D}FT (right).
`DfO' stands for density from orbitals,
`DF' for density functional,
`{\umlaut D}F' for double-dot-density functionals,
`EoM' for equation of motion.
The dashed line on the right indicates the possibility
of feeding the potentials with knowledge about $\vext$,
which gives rise to different realizations
of KS-TD{\umlaut D}FT.
}\label{flowchart}
\end{figure*}

%%%%%%%%%%%%%%%%%%%%%%%%%%%%%%%%%%%%%%%%%%%%%%%%%%%%%%%%%%%%%%%%%%%%%%%%%%%%%%%%%%%%%%%%%%%%%%%%%%%%

\section{KS-TDDFT}\label{secKSTDDFT}

In Sec. \ref{secDF} we formally defined a given observable as a functional of
$\ddn$. Now we want to apply the same idea to the KS approach, in an operational sense. The goal is to obtain
a closed set of equations that completely identifies the dynamics of the system
(see the left panel of Fig. \ref{flowchart}).

We start from the noninteracting initial-value problem
\be\label{iniphi}
\left\{\begin{array}{l}
i\partial_t\Phi(t)=\left(\hat{T}+\intdr \vs \rt \hat{n}({\bf r}) \right)\Phi(t)\\
\Phi(0)=\Phi_0.
\end{array}\right.
\ee
If the initial state $\Phi_0$ is such that
\be\label{ndnks}
\left\{
\begin{array}{l}
\bra{\Phi_0}\hat{n}\ket{\Phi_0}=\bra{\Psi_0}\hat{n}\ket{\Psi_0}\\
\bra{\Phi_0}\nabla \cdot \hat{{\bf j}}\ket{\Phi_0}=\bra{\Psi_0}\nabla \cdot \hat{{\bf j}}\ket{\Psi_0}%\\
%\partial^2_t \bra{\Psi(t)}\hat{n}\ket{\Psi(t)}=\partial^2_t \bra{\Phi(t)}\hat{n}\ket{\Phi(t)}.
\end{array}
\right.
\ee
and the potential is written as
\be\label{vs}
\vs\rt=\vext\rt+\vhxc \rt
\ee
with the Hartree-exchange-correlation (Hxc) potential $\vhxc \rt$ following from
\be\label{vhxc}
\nabla \cdot \big( n(t)\; \nabla \vhxc \rt \big)= q(t)-p(t),
\ee
where
$q(t)=\bra{\Psi(t)}\hat{q}({\bf r})\ket{\Psi(t)}$,
$p(t)=\bra{\Phi(t)}\hat{p}({\bf r})\ket{\Phi(t)}$, and
$\hat{p}({\bf r})$ is the noninteracting version of the operator $\hat{q}({\bf r})$,
then the density of the KS system, $n(t)=\bra{\Phi(t)}\hat{n}({\bf r})\ket{\Phi(t)}$,  matches the many-body density. This is
explicitly demonstrated in Appendix \ref{appn}.

Just like for $v$ in \eqref{ddfO}, if we add a gauge condition
we can formally solve Eq. \eqref{vhxc} as
\be\label{vhxcV}
\vhxc\rt=\mathcal{V}_{(\bf r)}[q(t)-p(t),n(t)].
\ee
Equation (\ref{vhxcV}) defines $\vhxc\rt$ as a functional of the KS and the many-body wave functions.
To transform it into a density functional,
we must write all expectation values in \eqref{vhxc} as density functionals.
Following Sec. \ref{secDF} [see Eq. (\ref{dfO})], the density-functional form of $q\rt$ is given by the set of equations
\be\label{dfq}
\left\{\begin{array}{l}
q\rt=\bra{\psi(t)}\hat{q}({\bf r})\ket{\psi(t)}\\
\\
i\partial_t\psi(t)=\left(\hat{T}+\hat{W}+\intdr v \rt \hat{n}({\bf r}) \right)\psi(t)\\
\psi(0)=\Psi_0\\
\\
n\rt=\bra{\psi(t)} \hat{n}({\bf r}) \ket{\psi(t)}\\
\mbox{gauge condition on }v\rt.
\end{array}\right.
\ee
Similarly, for the quantity $p\rt$ we have
\be\label{dfp}
\left\{\begin{array}{l}
p\rt=\bra{\phi(t)}\hat{p}({\bf r})\ket{\phi(t)}\\
\\
i\partial_t\phi(t)=\left(\hat{T}+\intdr w \rt \hat{n}({\bf r}) \right)\phi(t)\\
\phi(0)=\Phi_0\\
\\
n\rt=\bra{\phi(t)} \hat{n}({\bf r}) \ket{\phi(t)}\\
\mbox{gauge condition on }w\rt,
\end{array}\right.
\ee
where $\phi(t)$ and $w\rt$ are internal variables.

To close the circle, we need to connect the density appearing as input of the functionals
with the KS orbitals. However, using the expression $n\rt=\bra{\Phi(t)} \hat{n}({\bf r}) \ket{\Phi(t)}$
is not completely straightforward, as discussed in Ref. \onlinecite{Maitra2008}:
if implemented on a discrete time grid, the first nontrivial step, $t=\Delta t$, is consistently carried out
only if the density functionals are fed with $\dot{n}$
constructed via the continuity equation, namely
$\dot{n}({\bf r},\Delta t)
    =-\bra{\Phi(\Delta t)}\nabla \cdot \hat{{\bf j}}\ket{\Phi(\Delta t)}$.
This result can be generalized to all subsequent time steps, and we can write
\be\label{standardnphi}
%\dot{n}\rt\stackrel{\mbox{\scriptsize{?}}}{=}
\dot{n}\rt=
    -\bra{\Phi(t)}\nabla \cdot \hat{{\bf j}}\ket{\Phi(t)}.
\ee

In summary, Eqs.  \eqref{iniphi}--\eqref{standardnphi}
form a closed set that completely determines the evolution
of the KS system. This set involves the propagation of the many-body wave function,
hidden in one of the two density functionals needed to build $\vhxc$,
so from a practical point of view it is not more convenient than the original formulation
of the problem.
Moreover, constructing $q(\bfr,t)$ and $p(\bfr,t)$ as functionals of the density requires
the construction of $v$ and $w$, respectively, which makes the initial value problem
nonlocal in time.
Indeed, the very fact that KS-TDDFT can be consistently propagated in time is far from obvious:
it requires careful analysis\cite{Maitra2008} to demonstrate that the system is causal, i.e., it
does not require knowledge of the many-body density ahead of time.

\section{KS-TD\texorpdfstring{{\umlaut D}}{D}FT}\label{secKSTD"DFT}

Now let us go back to the idea of building a theory for $\ddot{n}$, rather than $n$.
Suppose that, in analogy with the standard KS approach,
we want to get $\ddot{n}$ from a fictitious, noninteracting system. However,
taking
\begin{equation}\label{ddotn_ddtPhi}
\ddot{n}\rt=\partial_t^2 \bra{\Phi(t)}\hat{n}\ket{\Phi(t)}
\end{equation}
as input into the Hxc functional would not work:
the system becomes non-predictive in the sense
that it has an infinity of solutions.
The reason is that the KS equation, being a first-order initial value problem, has one initial condition; but
Eq. (\ref{ddotn_ddtPhi}) involves second-order time derivatives, resulting in an under-determined problem.

A well-defined, time-local initial value problem is obtained as follows.
First, we raise the order of the KS equation
by transforming the first-order initial-value problem
\eqref{iniphi} into a (formally equivalent) second-order problem;
we then lower the order in $\ddot{n}$ by replacing Eq. (\ref{ddotn_ddtPhi}) with
\be\label{ddotn_j}
\ddot{n}\rt=-\partial_t\bra{\Phi(t)} \nabla\cdot\hat{{\bf j}}\ket{\Phi(t)},
\ee
which ensures that all explicit time-derivatives are at most of first order (see right panel of Fig. \ref{flowchart}).

Consider the first step in more detail. Simple time differentiation leads to
\be\label{soKS}
\begin{array}{c}
\left\{\begin{array}{l}
i\ddot{\Phi}(t)=\left(\hat{T}+\intdr \vs\rt\hat{n}({\bf r}) \right)\dot{\Phi}(t)+\\
                \hfill +\intdr \dvs\rt\hat{n}({\bf r})\Phi(t)\\
i\dot{\Phi}(0)=\left(\hat{T}+\intdr \vs({\bf r},0)\hat{n}({\bf r}) \right)\Phi_0\\
\Phi(0)=\Phi_0.
\end{array}\right.
\end{array}
\ee
Next, information about $\vext\rt$ needs to be fed into the system \eqref{soKS}.
Recall that in the first-order problem the information about the external potential
is passed to the KS system by writing $\vs\rt=\vext\rt+\vhxc \rt$
and density-functionalizing the Hxc part.
Now, if we write the second equation of \eqref{soKS} as
\be
i\dot{\Phi}(0)=\left(\hat{T}+\intdr (\vext({\bf r},0)+\vhxc({\bf r},0))\hat{n}({\bf r}) \right)\Phi_0,
\ee
where $\vhxc({\bf r},0)$ is given by \eqref{vhxc} at $t=0$, the system knows about $\vext({\bf r},0)$ and
then we only need to provide information about $\dvext\rt$.
This is achieved by writing the first equation  of (\ref{soKS}) as
\begin{eqnarray}\label{soeom}
i\ddot{\Phi}(t)&=&
\left(\hat{T}+\intdr v_1\rt\hat{n}({\bf r}) \right)\dot{\Phi}(t) \nonumber\\
&+&\intdr (\dvext\rt +v_2\rt)\hat{n}({\bf r})\Phi(t)
\end{eqnarray}
and density-functionalizing  $v_1\rt$ and $v_2\rt$.
We proceed along similar lines as
in  Sec. \ref{secKSTDDFT}, ensuring that the KS $\ddn$ matches the many-body $\ddn$.
In detail:
\be\label{v1v2}\begin{array}{l}
v_1\rt=v\rt+\Vr[q(t)-p(t),n(t)]\\
v_2\rt=\Vr[q_1(t)+v(t) q_2(t)-p_1(t)-v(t) p_2(t)+\\
    \;\;\;\;-\nabla \cdot \big( -\nabla\cdot {\bf j}\rt\; \nabla \Vr[q(t)-p(t),n(t)]\big),n(t)]
\end{array}\ee
with
\be\begin{array}{l}
v\rt=\Vr[\ddn(t)-q(t),n(t)]\\
q(t)=\bra{\psi(t)}\hat{q}({\bf r})\ket{\psi(t)}\\
p(t)=\bra{\phi(t)}\hat{p}({\bf r})\ket{\phi(t)}\\
n(t)=\bra{\phi(t)}\hat{n}({\bf r})\ket{\phi(t)},
\end{array}\ee
\be\begin{array}{l}
q_1\rt=\bra{\psi(t)}i[\hat{T}+\hat{W},\hat{q}({\bf r})]\ket{\psi(t)}\\
q_2\rt=\bra{\psi(t)}i[\hat{n},\hat{q}({\bf r})]\ket{\psi(t)}\\
p_1\rt=\bra{\phi(t)}i[\hat{T},\hat{p}({\bf r})]\ket{\phi(t)}\\
p_2\rt=\bra{\phi(t)}i[\hat{n},\hat{p}({\bf r})]\ket{\phi(t)}\\
\nabla\cdot {\bf j}\rt=-\bra{\phi(t)}i[\hat{T},\hat{n}]\ket{\phi(t)},
\end{array}\ee
and
\be\label{psiphi}\begin{array}{l}
i\partial_t\psi(t)=\Big(\hat{T}+\hat{W}+\intdr \! \Vr[\ddn(t)-q(t),n(t)] \hat{n}({\bf r}) \Big)\psi(t)\\
\psi(0)=\Psi_0\\
\\
i\partial_t\phi(t)=\Big(\hat{T}+\intdr \Vr[\ddn(t)-p(t),n(t)] \hat{n}({\bf r}) \Big)\phi(t)\\
\phi(0)=\Phi_0.
\end{array}\ee
Notice that now $\ddn$ is the sole input variable, while $n$ is treated as an internal variable.
This set of equations completely defines the functionals required by TD{\umlaut D}FT.
Our realization of the functionals entails the propagation
of two wave functions, a many-body $\psi$ and a single-particle  $\phi$,
which store the memory of the functional about past values of $\ddn$.
This yields the potentials ``on the fly'' during the propagation
of the equations of motion. In other words, to propagate $v_1(t)$
and $v_2(t)$ the functional only uses $\psi(t)$
and $\phi(t)$ at time $t$, which encodes the values $\{\ddot{n}(0\leq t'< t)\}$.
Thus, by propagating two auxiliary wave functions alongside the KS wave function $\Phi(t)$,
the functional technically becomes local in time while still retaining its memory of past densities.

A central task of the constituting set of equations is to make sure that the external potential is appropriately accounted for.
It turns out that in doing so we have a choice, which does not affect the exact solution,
but may have consequences for approximations. In \eqref{v1v2} we have made what we
call the \emph{minimal} choice (m-KS), replacing $v_{\rm ext}$ with the internal variable $v$ everywhere.
By contrast, the \emph{maximal} choice (M-KS) would be to explicitly keep $v_{\rm ext}$
in both equations of \eqref{v1v2}. Mixed choices are possible as well, such as
feeding $\vext$ to $v_1$, but not $v_2$ (Mm-KS), or vice versa (mM-KS).
These choices are equivalent in the sense that they lead to the same solution of the system.
In deriving approximations, however, the particular choice of approach may make a substantial difference.

Carefully distinguishing between these choices allows one to establish the
connection between KS-TD{\umlaut D}FT and KS-TDDFT: one finds
that KS-TDDFT is equivalent to Mm-KS-TD{\umlaut D}FT.
The two potentials $v_1$ and $v_2$ then become related via $v_2\rt=\partial_t(v_1\rt-\vext\rt)$,
reducing the two degrees of freedom to one, the Hxc potential of Eq. \eqref{vhxcV},
which satisfies $\vhxc\rt= v_1\rt-\vext\rt$.

This formal result provides important insight into the nature of KS-TDDFT:
it clarifies the way the KS-TDDFT system, when combined with the exact Hxc functional,
propagates in time. KS-TD{\umlaut D}FT, being a second-order differential problem using $\ddot n$ as basic variable, is a time-local and demonstrably causal
initial-value problem, and KS-TDDFT inherits this property.
Furthermore, we can state more precisely what the ``memory'' of the Hxc functional actually is,
namely the fact that $\dot{v}_{\rm Hxc}(t)$ is uniquely defined
by the initial states and the set $\{\ddot{n}(0\leq t'< t)\}$.

%%%%%%%%%%%%%%%%%%%%%%%%%%%%%%%%%%%%%%%%%%%%%%%%%%%%%%%%%%%%%%%%%%%%%%%%%%%%%%%%%%%%%%%%%%%%%%
%%%%%%%%%%%%%%%%%%%%%%%%%%%%%%%%%%%%%%%%%%%%%%%%%%%%%%%%%%%%%%%%%%%%%%%%%%%%%%%%%%%%%%%%%%%%%%
\section{Approximations}\label{secAppr}

The complete set of equations for KS-TD{\umlaut D}FT, Eqs. (\ref{ddotn_j}--(\ref{psiphi}), in any of its realizations,
can be used as starting point for further developments aimed
at simplifying their structure before intervening with approximations.
Some simple approximations, however, can readily be applied to the system
in its actual form.

\subsection{Exchange-only and beyond}

The characterization of density-functionals provided at the end
of Section \ref{secDF} involves a wave function that comes from
a many-body problem which is more complex than
the usual Schr\"odinger equation, involving the inverse Sturm-Liouville operator;
likewise for the set of KS-TD{\umlaut D}FT equations.
More specifically, the set involves the propagation
of one many-body state, $\psi$,
and two single-particle states, $\phi$ and $\Phi$.
The states $\psi$ or $\phi$ are used
to evaluate the expectation values of certain operators which,
in turn, are used to define the potentials $v_1$ and $v_2$ that
enter in the equation of motion for $\Phi$.

The first, simplest approximation is to take all expectation values
appearing in this scheme with respect to the KS wave function $\Phi$, instead of $\phi$ and $\psi$.
This has the twofold advantage of avoiding to deal with a many-body
wave function and to propagate more than one equation of motion.
In the case of Mm-KS-TD{\umlaut D}FT (the version equivalent to KS-TDDFT),
one finds that this reduces to
\begin{equation}\label{LHXO}
\nabla \cdot [n(\bfr,t) \nabla v_{\rm Hxc}(\bfr,t)] \approx
\langle \Phi(t) | \partial_\nu \hat W_{\nu}(\bfr)|\Phi(t)\rangle .
\end{equation}
This expression was introduced earlier by Ruggenthaler and Bauer, \cite{Ruggenthaler2009}
who called it the local Hartree-exchange-only (LHXO) approximation.
Working out the right-hand side of Eq. (\ref{LHXO}) gives following explicit approximation for the LHXO exchange potential
(assuming that the system is not spin polarized):
\begin{equation}\label{LHXO_explicit}
\nabla v_{\rm x}^{\rm LHOX}(\bfr,t)
= - \int d\bfr' \frac{|\gamma(\bfr,\bfr',t)|^2}{n(\bfr,t)}\left(\nabla \frac{1}{|\bfr - \bfr'|}\right),
\end{equation}
where $\gamma(\bfr,\bfr')$ is the KS one-particle reduced density matrix. Equation (\ref{LHXO_explicit}) can be
compared with the well-known Slater exchange potential:\cite{Slater1951,GiulianiVignale}
\begin{equation}\label{Slater}
v^{\rm Slater}_{\rm x}(\bfr,t)
= - \int d\bfr' \frac{|\gamma(\bfr,\bfr',t)|^2}{n(\bfr,t)|\bfr-\bfr'|}.
\end{equation}
The Slater potential (which is derived in a different way) has been widely studied in the literature, and is known to be a decent
approximation to the exact exchange potential.\cite{Krieger1992}
The performance of the LHXO potential, although much less studied, seems to be comparable to that of the Slater potential.\cite{Ruggenthaler2009}

The approximation (\ref{LHXO}) can be improved in several ways.
A systematic approach to constructing more and more accurate approximations is via the
G\"orling-Levy perturbation theory,\cite{Goerling1994,Goerling1997,Goerling2006}
according to which the many-body wave function $\Psi(t)$ is  expanded as
\begin{equation}\label{GL}
\Psi(t) = \Phi(t) + \Psi^{(1)}(t) + \Psi^{(2)}(t) + \ldots .
\end{equation}
This expansion is in orders of a coupling parameter $\lambda$, whose significance
is discussed in Appendix \ref{xxtddft}. To first order in $\lambda$, G\"orling-Levy perturbation theory
defines the exact-exchange formulation of TDDFT.\cite{Goerling1997,Goerling2006}
We show in Appendix \ref{xxtddft} that by replacing $\psi$ in the  Mm-KS-TD{\umlaut D}FT scheme
with the KS wave function $\Phi(t)$ plus its first-order correction $\Psi^{(1)}(t)$,
we indeed recover the exact exchange potential of KS-TDDFT.

The flexibility of the formalism of Sec. \ref{secKSTD"DFT} can be exploited for constructing
additional approximations at the exchange-only level. In m-KS-TD{\umlaut D}FT, the ``internal'' many-body
wave function $\psi$ appears in several places: $q$, $q_1$, and $q_2$, which in turn enter into $v_1$ and $v_2$.
At each occurrence, $\psi$ can be approximated by $\Phi$ or $\Phi + \Psi^{(1)}$.
The various possible combinations then give rise to approximations that are intermediate between LHXO
and exact exchange.

Higher orders of the $\lambda$-expansion can in principle be used for a systematic treatment of time-dependent correlation,
thus extending the KS-TD{\umlaut D}FT formalism beyond exact exchange.

\subsection{Adiabatic approximation}

The adiabatic approximation plays a central role in TDDFT, at the formal level as well as in practice.
We will first discuss the adiabatic approximation in more general terms, and then in the specific context of KS-TD{\umlaut D}FT.

The adiabatic theorem \cite{Born1928,Kato1950} states that,
for sufficiently slowly varying time-dependent Hamiltonians, a many-body system which at the initial time $t_0$
is in an eigenstate of the Hamiltonian at $t_0$, will remain, in good approximation,
proportional to the instantaneous eigenstate of the nonstationary Hamiltonian.
More precisely, if $\psi(t_0)=\varphi_n^{t_0}$
with $\hat{H}(t_0)\varphi_n^{t_0}=E_n^{t_0}\varphi_n^{t_0}$,
then for $t\ge t_0$ the solution to the time-dependent Schr\"odinger equation
$i\partial_t\psi(t)=\hat{H}(t)\psi(t)$ can be approximated  as\cite{Chruscinski2004}
\be \label{adiabatic_theorem}
\psi(t)=e^{i\xi_n(t)}e^{-i\int_{t_0}^t E_n^\tau d\tau}\varphi_n^{t},
\ee
with $\hat{H}(t)\varphi_n^{t}=E_n^{t}\varphi_n^{t}$ and the phase $\xi_n(t)$ following from
$\partial_t \xi_n(t)=i\braket{\varphi_n^{t}|\partial_t \varphi_n^{t}}$.

In the context of TDDFT, the term ``adiabatic approximation'' requires some clarification.\cite{Ullrich2012}
It is most commonly used in situations where the KS time propagation is carried out with an approximate ground-state
xc potential, such as the local-density approximation (LDA), evaluated at the time-dependent density at the same time $t$:
\begin{equation}
v_{\rm xc}^{\rm ad,app}(\bfr,t) = \left. v_{\rm xc}^{\rm gs,app}[n_{\rm gs}](\bfr)\right|_{n_{\rm gs}(\bfr)\to n(\bfr,t)},
\end{equation}
where $n_{\rm gs}(\bfr)$ is a ground-state density. The vast majority of applications of KS-TDDFT are performed in this way.

If the approximate ground-state xc potential $v_{\rm xc}^{\rm gs,app}$ is an explicit density functional,
then the adiabatic approximation is straightforward. However, many approximate xc functionals are given in terms of the KS orbitals at time $t$,
such as the Slater potential (\ref{Slater}) or the KLI potential.\cite{Krieger1992} The KS orbitals are implicit density functionals:
via the KS equation they carry a memory of past densities. Therefore, the Slater potential is an adiabatic {\em orbital} functional, but not an
adiabatic {\em density} functional.

A procedure to construct adiabatic approximations to explicit orbital functionals was given in Ref. \onlinecite{Wijewardane2008} for the case of exact exchange.
It consists of two steps, to be carried out at each step of the time propagation:
(1) Find a noninteracting system which produces $n(\bfr,t)$ as its ground-state density. There are many numerical schemes
which can do this efficiently.\cite{Jensen2017} (2) Use the resulting static KS orbitals (and eigenvalues, if needed) to evaluate the xc orbital functional
at time $t$.

Lastly, the {\em adiabatically exact} KS potential
is defined as that local potential $v^t_{\rm loc}(\bfr)$ which yields the exact time-dependent density $n(\bfr,t)$ at time $t$
as its ground-state density, in a noninteracting system. \cite{Thiele2008}
The adiabatically exact xc potential can be constructed as
\begin{equation}
 v_{\rm xc}^{\rm ad,exact}(\bfr,t) = v^t_{\rm loc}(\bfr) - v_{\rm ext}^t(\bfr)
- \int d\bfr' \frac{n(\bfr',t)}{|\bfr-\bfr'|} \:.
\end{equation}
Here, $v_{\rm ext}^t(\bfr)$ is the external one-body potential which produces $n(\bfr,t)$ as the ground-state density
in an interacting many-body system. To obtain it explicitly requires inversion of the static many-body Schr\"odinger equation, which
is much harder than inverting the KS equation.\cite{Coe2009}

Compared to this, the definition of an adiabatically exact scheme is more straightforward in KS-TD{\umlaut D}FT.
All we need to do is to replace the full time propagation of the states
$\psi$ and $\phi$,  Eqs. (\ref{psiphi}), with the instantaneous eigenstate problems
\begin{eqnarray}
\left(\hat{T}+\hat{W}
+\intdr \Vr[\ddn(t)-q_\psi^t,n_\psi^t] \hat{n}({\bf r}) \right)\psi^t
&=&  E_\psi^t\psi^t,
\nonumber\\
\left(\hat{T} +\intdr \Vr[\ddn(t)-p_\phi^t,n_\phi^t]
\hat{n}({\bf r}) \right)\phi^t
&=& E_\phi^t\phi^t, \nonumber \\
\end{eqnarray}
where $q_\psi^t = \langle \psi^t | \hat q | \psi^t\rangle$, $n_\psi^t = \langle \psi^t | \hat n | \psi^t\rangle$,
$p_\phi^t = \langle \phi^t | \hat q | \phi^t\rangle$, and $n_\phi^t = \langle \phi^t | \hat n | \phi^t\rangle$.
This prescription is compatible with the adiabatic limit of the wave functions, Eq. (\ref{adiabatic_theorem}).
Notice, in particular, that the inversion of the many-body Schr\"odinger equation is avoided, and,
when applied to Mm-KS-TD{\umlaut D}FT, the corresponding exact adiabatic Hxc potential
is obtained as
\begin{equation}
v_{\rm Hxc}^{\rm ad,exact}(\bfr,t) =  \Vr[q_\psi^t -p_\phi^t,n_\phi^t] .
\end{equation}

%%%%%%%%%%%%%%%%%%%%%%%%%%%%%%%%%%%%%%%%%%%%%%%%%%%%%%%%%%%%%%%%%%%%%%%%%%%%%%%%%%%%%%%%%%%%%%
%%%%%%%%%%%%%%%%%%%%%%%%%%%%%%%%%%%%%%%%%%%%%%%%%%%%%%%%%%%%%%%%%%%%%%%%%%%%%%%%%%%%%%%%%%%%%%

\section{Conclusions} \label{secConclusions}

Our point of departure in this paper was the general statement that in TDDFT the many-body wave
function and all properties derivable from it are, in principle, functionals of the time-dependent density $n(\bfr,t)$
and the initial state. But what do these functionals look like? In Ref. \onlinecite{Ullrich2012}, the universal density functional
of the exchange-correlation potential was characterized as being similar to a preexisting library, where one can simply look up the potential
for any given density (and initial state). In this paper, we adopt a slightly different perspective and view density functionals as self-contained
procedures with the many-body problem built into the internal machinery, but hidden from sight.

Based on this definition of functionals, KS theory at the exact level can be formulated as a closed procedure.
The result is an initial value problem that has a more complicated structure than the usual many-body Schr\"odinger
equation, explicitly involving the previous history of the system. Demonstrating the causality of KS-TDDFT,
especially right around the initial time, is a somewhat subtle affair.\cite{Maitra2008}

It turns out, however, that KS-TDDFT can be reformulated in such a way that it involves internally a much simpler, explicitly time-local
initial value problem. All one needs to do is to adopt the second time derivative of the density, $\ddot n(\bfr,t)$, as basic variable,
which is featured in the force-balance equation of TDDFT. The resulting KS-TD{\umlaut D}FT formalism, which is of second order in time,
offers a simpler point of view of the question of causality.
This might provide a new way forward in the ongoing attempts to solidify the foundations of TDDFT regarding existence, uniqueness
and $v$-representability.\cite{Ruggenthaler2015}

On the practical side, KS-TD{\umlaut D}FT could be useful for constructing new approximations to the exchange-correlation potential.
Here, we have discussed in detail the exchange-only case using a perturbative treatment, and we have shown how to construct the
exact adiabatic approximation. These ideas, if developed further, could be the basis for new approximations
to capture nonadiabatic correlation effects.
Related efforts to further illustrate KS-TD{\umlaut D}FT are currently underway using simple model systems.

%\begin{acknowledgements}
\acknowledgments
C.U. acknowledges support by NSF grant No. DMR-1810922.
%\end{acknowledgements}
\section*{AIP Publishing data sharing policy}
Data sharing is not applicable to this article 
as no new data were created or analyzed in this study.

\bibliography{shortbib.bib}
\appendix

\section*{Appendices}

\subsection{The \texorpdfstring{$\mathcal{V}_{(\bf r)}$}{V(r)} operator}\label{appV}

In going from Eq. \eqref{ddfO} to Eq. \eqref{ddVfO} we have assumed
that the last relation of \eqref{ddfO},
complemented with a suitable gauge condition, allows us to completely identify $v\rt$,
given all other functions. Written in a more compact way, Eq. \eqref{ddfO} implies
\be\label{sturm}
\nabla \cdot ( f\rt \nabla v \rt)=g\rt ,
\ee
where $f\rt$ and $g\rt$ are given real functions. The formal solution of this is
\be
v\rt=\Vr[g(t),f(t)].
\ee
As first recognized by van Leeuwen,\cite{vanLeeuwen1999}
Eq. (\ref{sturm}) requires the inversion of the Sturm-Liouville operator
$\nabla \cdot ( f\rt \nabla \ldots)$,
which is indeed well defined when supplemented with a gauge condition.
Unfortunately, for generic $f$ and $g$
there is no explicit expression for this inverse operator,
but efficient numerical methods to solve Eq. \eqref{sturm} are available.

Perhaps a more intuitive perspective on Eq. \eqref{sturm}
can be gained from the observation that in the case
of homogeneous density [$\nabla n\rt=0$, or,
in the notation of  Eq. \eqref{sturm}, $\nabla f\rt=0$],
it has the form of Poisson's equation, the  cornerstone of electrostatics:
\be\label{poisson}
\nabla^2 \phi\rt=-\frac{\rho\rt}{\varepsilon},
\ee
$\rho\rt$ being the charge distribution and $\varepsilon$ the dielectric constant,
which has the analytic solution
\be\label{phisol}
\phi\rt=\intdrr \frac{\rho\rrt}{ 4 \pi \varepsilon|{\bf r}'-{\bf r}|}.
\ee
In fact, Eq. \eqref{sturm} can be regarded as Poisson's equation
in an inhomogeneous medium.
More specifically, for a dielectric function of the form
$\varepsilon\rt$, Poisson's equation reads
\be\label{poissoninh}
\nabla \cdot ( \varepsilon\rt \nabla \phi \rt)=-\rho\rt ,
\ee
which has the exact form of Eq. \eqref{sturm};
also notice that neither $\varepsilon\rt$ nor $n\rt$ can take negative values.

Numerical solution of the inhomogeneous Poisson equation, Eq. \eqref{poissoninh},
and hence of Eq. \eqref{sturm}, can be achieved by iteration:
\begin{multline}
v\rt=\intdrr\frac{1}{ 4 \pi |{\bf r}'-{\bf r}|}\left(
\frac{\ddn \rrt-q\rrt}{ n\rrt}+\right.\\
\left.+\frac{\nabla' n\rrt}{ n\rrt} \cdot \nabla'v\rrt
\right),
\end{multline}
where the function $v$ appears on both sides.
If we start iterating by setting $v\rt=0$ on the right-hand side,
the first iteration is exact in the case $\nabla n\rt=0$,
while subsequent iterations contain higher and higher spatial derivatives of the density.
In other words, the solution is built around a
homogeneous-electron-gas--like approximation
by adding terms more and more sensitive to inhomogeneities of the density.

\subsection{Equivalence of many-body and KS densities}\label{appn}

For the many-body (MB) system we have
\be\label{mbq}
\ddot{n}_{\rm MB}\rt=\nabla\cdot \left( n_{\rm MB}\rt \nabla \vext\rt\right)+\bra{\Psi(t)}\hat{q}\ket{\Psi(t)},
\ee
where $n_{\rm MB}\rt=\bra{\Psi(t)}\hat{n}\ket{\Psi(t)}$. The corresponding relation
for the KS system is given by
\be
\ddot{n}_{\rm KS}\rt=\nabla\cdot \left( n_{\rm KS}\rt \nabla \vs\rt\right)
+\bra{\Phi(t)}\hat{p}\ket{\Phi(t)},
\ee
where $n_{\rm KS}\rt=\bra{\Phi(t)}\hat{n}\ket{\Phi(t)}$.
Using Eqs. \eqref{vs} and \eqref{vhxc} we can rewrite the latter as
\be\label{ksq}
\ddot{n}_{\rm KS}\rt=\nabla\cdot \left( n_{\rm KS}\rt \nabla \vext\rt\right)
+\bra{\Psi(t)}\hat{q}\ket{\Psi(t)}.
\ee
By subtracting Eq. \eqref{ksq} from Eq. \eqref{mbq} and introducing the function
$\delta \rt=n_{\rm MB}\rt-n_{\rm KS}\rt$, we obtain
\be\label{dddelta}
\ddot{\delta}\rt=\nabla\cdot \left( \delta \rt \nabla \vext\rt\right).
\ee
From the initial conditions \eqref{ndnks} it follows that
\be\label{deltaini}
\delta({\bf r},0)=0 ,\qquad
\dot{\delta}({\bf r},0)=0.
\ee
Together, Eqs. \eqref{deltaini} and \eqref{dddelta} form a well-defined
initial value problem with a unique solution.
By direct check one can see that $\delta\rt=0$ is indeed a solution,
and hence $n_{\rm MB}\rt=n_{\rm KS}\rt$.

\subsection{Derivation of exact-exchange TDDFT}\label{xxtddft}

We begin with the equation of motion of the density,
\begin{equation}\label{eom}
\ddot n(\bfr,t) = \nabla \cdot [n(\bfr,t) \nabla v(\bfr,t)] + q(\bfr,t),
\end{equation}
see Eq. (\ref{7}), where
\begin{equation}
q(\bfr,t) = \langle \Psi(t) | \partial_\nu \partial _\mu \hat T_{\mu\nu}(\bfr)+ \partial_\mu \hat W_{\mu}(\bfr)|\Psi(t)\rangle \:.
\end{equation}
Written in second-quantized notation, the momentum-stress tensor is
\begin{eqnarray}
\hat T_{\mu\nu}(\bfr) &=& \frac{1}{2} \sum_\sigma \bigg\{ [\partial_\nu \hat\psi_\sigma^\dagger(\bfr)]\partial_\mu \hat\psi_\sigma(\bfr)
+ [\partial_\mu \hat\psi_\sigma^\dagger(\bfr)]\partial_\nu \hat\psi_\sigma(\bfr) \nonumber\\
&&
{}- \frac{1}{2}\partial_\mu \partial_\nu [ \hat\psi_\sigma^\dagger(\bfr)\hat\psi_\sigma(\bfr)]
\bigg\}
\end{eqnarray}
and the divergence of the interaction-stress tensor is
\begin{equation}
\hat W_{\nu}(\bfr)= \sum_{\sigma\sigma'}\int \! d\bfr'\!\left[\partial_\nu \frac{1}{|\bfr-\bfr'|}\right]\hat\psi_\sigma^\dagger(\bfr)\hat\psi_{\sigma'}^\dagger(\bfr')
\hat\psi_{\sigma'}(\bfr')\hat\psi_\sigma(\bfr) \:.
\end{equation}
Next, we use G\"orling-Levy perturbation theory \cite{Goerling1994,Goerling1997,Goerling2006} to connect
Eq. (\ref{eom}) with the time-dependent optimized effective potential (TDOEP) forma\-lism; \cite{Ullrich1995} specifically,
we will derive the limit of exact exchange.

We begin by defining a modified time-dependent Schr\"odinger equation as follows:
\begin{equation}
i\frac{\partial }{\partial t} \Psi^\lambda(t) = \left[ \hat{T} + \hat{V}^\lambda(t) + \lambda \hat{W}\right] \Psi^\lambda(t) \:.
\end{equation}
The single-particle potential $v^\lambda(\bfr,t)$ is chosen
such that the density is the same for each $\lambda$ along the so-called adiabatic connection, $0\le \lambda \le 1$:
\begin{equation}
\langle \Psi^\lambda(t) | \hat{n}(\bfr) | \Psi^\lambda(t)\rangle = n(\bfr,t) \:,
\end{equation}
where $v^{\lambda=1}(\bfr,t)=v(\bfr,t)$ is the given external potential, and $v^{\lambda=0}(\bfr,t)=v_s(\bfr,t)$
is the time-dependent KS potential. Next, $v^\lambda(\bfr,t)$ is expanded as
\begin{equation} \label{13.vlambdaTaylor}
v^\lambda(\bfr,t) = v_s(\bfr,t) + \lambda v^{(1)}(\bfr,t) + \lambda^2 v^{(2)}(\bfr,t) + \ldots \:.
\end{equation}
From the definition of the time-dependent KS potential, one immediately finds that
\begin{equation}
v_{\rm H}(\bfr,t) + v_{\rm xc}(\bfr,t) = -v^{(1)}(\bfr,t) - v^{(2)}(\bfr,t) - \ldots \:.
\end{equation}
The first-order term is the sum of the negative Hartree and local exchange potentials:
\begin{equation}\label{v1}
v^{(1)}(\bfr,t) = -v_{\rm H}(\bfr,t) - v_{\rm x}(\bfr,t) \:.
\end{equation}
All higher-order terms taken together yield the correlation potential: $v^{(2)}+ v^{(3)}+...= -v_{\rm c}$\:.

The time-dependent many-body wave function can also be expanded in powers of $\lambda$.
To zeroth order in $\lambda$, one finds $\Psi^{(0)}(t)=\Phi(t)$ (the KS wave function).
The first-order correction is given by\cite{Goerling1997,Goerling2006}
\begin{eqnarray}\label{13.Psifirst}
\lefteqn{\Psi^{(1)}(t)=}\\
&&
 \frac{1}{i}\sum_{k=0}^\infty \int^t dt' \: \Phi_k(t) \langle \Phi_k(t') | \hat{W} - \hat{V}_{\rm H}(t')
- \hat{V}_{\rm x}(t')|\Phi_0(t')\rangle. \nonumber
\end{eqnarray}
The $\Phi_k(t)$ are time-dependent Slater determinants, built using occupied and unoccupied orbitals that follow from
time propagation under the influence of the time-depen\-dent KS Hamiltonian $\hat{H}_s(t)$.

Let us now return to Eq. (\ref{eom}), generalizing it along the adiabatic connection:
\begin{eqnarray}\label{eom_lambda}
\ddot n(\bfr,t) &=& \nabla \cdot [n(\bfr,t) \nabla v^\lambda(\bfr,t)] + \langle \Psi^\lambda(t) | \partial_\nu \partial _\mu \hat T_{\mu\nu}(\bfr)|\Psi^\lambda(t)\rangle
\nonumber\\
&+&
\lambda \langle \Psi^\lambda(t) | \partial_\nu \hat W_{\nu}(\bfr)|\Psi^\lambda(t)\rangle.
\end{eqnarray}
The zeroth-order equation,
\begin{equation}\label{eom_0}
\ddot n(\bfr,t) = \nabla \cdot [n(\bfr,t) \nabla v_s(\bfr,t)] + \langle \Phi_0(t) | \partial_\nu \partial _\mu \hat T_{\mu\nu}(\bfr)|\Phi_0(t)\rangle,
\end{equation}
simply expresses the fact that the noninteracting KS system produces the exact density.
The first-order equation reads
\begin{eqnarray}\label{eom_1}
0 &=& \nabla \cdot [n(\bfr,t) \nabla v^{(1)}(\bfr,t)] + \langle \Psi^{(1)}(t) | \partial_\nu \partial _\mu \hat T_{\mu\nu}(\bfr)|\Phi_0(t)\rangle
\nonumber\\
&+&
\langle \Phi_0(t) | \partial_\nu \partial _\mu \hat T_{\mu\nu}(\bfr)|\Psi^{(1)}(t)\rangle
+\langle \Phi_0(t) | \partial_\nu \hat W_{\nu}(\bfr)|\Phi_0(t)\rangle\nonumber\\
\end{eqnarray}
which gives, using Eq. (\ref{v1}),
\begin{eqnarray}\label{eom_1a}
\lefteqn{
\nabla \cdot [n(\bfr,t) \nabla v_{\rm Hx}(\bfr,t)] = \langle \Psi^{(1)}(t) | \partial_\nu \partial _\mu \hat T_{\mu\nu}(\bfr)|\Phi_0(t)\rangle}
\nonumber\\
&+&\langle \Phi_0(t) | \partial_\nu \partial _\mu \hat T_{\mu\nu}(\bfr)|\Psi^{(1)}(t)\rangle+
\langle \Phi_0(t) | \partial_\nu \hat W_{\nu}(\bfr)|\Phi_0(t)\rangle \:.\nonumber\\
\end{eqnarray}
Equation (\ref{eom_1a}) is different from the local Hartree-ex\-change-only (LHXO) approximation
of Ruggenthaler and Bauer, \cite{Ruggenthaler2009} see Eq. (\ref{LHXO}), due to the presence of the terms featuring $\Psi^{(1)}(t)$.
Instead, Eq. (\ref{eom_1a}) defines the  {\em exact} exchange potential of TDDFT, $v_{\rm x}(\bfr,t)$.
To see this, one needs to prove that Eq. (\ref{eom_1a}) is equivalent to the exact-exchange TDOEP equation. \cite{Ullrich1995}
Indeed, Liao {\em et al.} \cite{Liao2017,Liao2018} have shown that the TDOEP 
equation---an integral equation over time\cite{Mundt2006,Wijewardane2008}---can be recast into a time-local differential
equation featuring $\nabla \cdot [n(\bfr,t) \nabla v_{\rm x}(\bfr,t)]$. It is a technically straightforward though somewhat tedious exercise to
show that the right-hand side of Eq. (\ref{eom_1a}) reduces to the right-hand side of Eq. (20) of Ref. \onlinecite{Liao2017}, thus establishing
the desired equivalence.

\end{document}